\documentclass[twocolumn,showpacs,showkeys,prd]{revtex4}
\usepackage{graphicx}

\def\GeV{\hbox{ GeV}}

\begin{document}
%%%%%%%%%%%%%%%%%%%%%%%%%%%%%%%%%%%%%%%%%%%%%%%%%%%%%%%%%%%%%%%%%%%%%%

\title{Lightest Higgs boson masses \\in the $R$-parity violating
  supersymmetry}

\author{Marek G\'o\'zd\'z}
\email{mgozdz@kft.umcs.lublin.pl}

\affiliation{
Department of Informatics, Maria Curie-Sk{\l}odowska University, \\
pl.~Marii Curie--Sk{\l}odowskiej 5, 20-031 Lublin, Poland}

\begin{abstract}
  The first results on the searches of the Higgs boson appeared this
  Summer from the LHC and Tevatron groups, and has been recently backed
  up by the ATLAS and CMS experiments taking data at CERN's LHC. Even
  though the excitement that this particle has been detected is still
  premature, the new data constrain the mass of the lightest Higgs boson
  $m_{h^0}$ to a~very narrow 120--140~GeV region with a~possible peak at
  approximately 125 GeV.

  In this communication we shortly present the Higgs sector in a~minimal
  supergravity model with broken $R$-parity. Imposing the constraint on
  $m_{h^0}$ we show that there is a~relatively large set of free
  parameters of the model, for which that constraint is fulfilled. We
  indetify also points which result in the lightest Higgs boson mass
  being approximately 125 GeV. Also the dependence on the magnitude of
  the $R$--parity admixture to the model is discussed. 
\end{abstract}


\maketitle

%%%%%%%%%%%%%%%%%%%%%%%%%%%%%%%%%%%%%%%%%%%%%%%%%%%%%%%%%%%%%%%%%%%%

\section{Introduction}

The Standard Model and most of its supersymmetric extensions suffer from
being a~theory of massless particles. Therefore a~mechanism that would
provide masses is required. Among several possibilities the Higgs
mechanism plays a~major role. It assumes the existence of an additional
scalar field, the Higgs field, which has non-zero vacuum expectation
value (vev). The correct implementation of this mechanism leads in the
Standard Model not only to massive gauge bosons (with the photon
correctly remaining massless), but also to massive fermions, and a
proper electroweak symmetry breaking from the weak gauge groups $SU(2)_L
\times U(1)_Y$ to the electromagnetism $U(1)_q$. These features make
this mechanism an extremely convenient and elegant solution. The
experimental smoking gun confirming this theory would be the discovery
of the Higgs boson.

Earlier this year the Tevatron collaborations CDF and D0 reported an
excess of events in the Higgs to two photons channel ($H\to
\gamma\gamma$) observed in the mass region 120--140 Gev
\cite{Tevatron}. The significance of these data were reported on the
level of $2.5\sigma$. Only recently the newly announced LHC--7 results
\cite{lhc7} from the ATLAS and CMS Collaborations confirmed an excess of
events in the same channel within 115--130 GeV range, with a~maximum at
125 GeV, at the statistical significance of $\approx 2\sigma$. Even
though $2.5\sigma$ cannot be named a~discovery, an effect independently
obtained within very similar mass ranges by four project working on two
biggest accelerators in the world may give hope that some new particle
has been observed.

Even though the experimental results are at first interpreted within the
standard model, its supersymmetric and other exotic extensions of
various kinds can also be tested against the newly reported findings
\cite{other}. Following this line of research, in this communication we
discuss an $R$-parity violating minimal supergravity (RpV mSUGRA) model
and constrain it by the liberal condition for the lightest Higgs boson
mass $120 \GeV \le m_{h^0} \le 140 \GeV$. Also the specific case of
$m_{h^0} \approx 125\GeV$ is considered.

\section{The model}

The $R$-parity is a~multiplicative quantum number defined as
$R=(-1)^{2s+3(b-\ell)}$ and implies stability of the lightest
supersymmetric particle. In this paper, following Ref.~\cite{mSUGRA}, we
adopt the so-called generalized baryon parity in the form of a~discrete
$Z_3$ symmetry $B_3=R_3L_3$ which ensures the stability of the proton
and lack of unwanted dimension-5 operators. In short, $R$ is equal to +1
for ordinary particles, and $R=-1$ for supersymmetric partners, and this
$R$ is usually assumed to be conserved in interactions. This assumption,
however, is based mainly on our will to exclude lepton and baryon number
violating processes, which has not been observed in the low-energy
regime. Notice, that the generation lepton numbers $\ell_e$, $\ell_\mu$,
and $\ell_\tau$, also conserved in the standard model, are broken in the
neutrino oscillations. There is in fact no underlying principle which
forbids breaking of $\ell$ or $b$. The baryon number violation is highly
constrained by the proton decay, but the lepton number violation may
occur at high energies. In general supersymmetric models one often
discusses the possibility of having the $R$-parity violating terms,
properly suppressed, in the theory. These are the so-called $R$--parity
violating models.

We perform the calculations within the framework described in detail in
Ref.~\cite{mSUGRA}. This model takes into account full dependence of the
mass matrices and renormalization group equations on the $R$--parity
violating couplings. We define it below by quoting the expressions for
the superpotential and soft supersymmetry breaking Lagrangian. Next, we
discuss the free parameters of the model, and the Higgs sector.

The interactions are defined by the superpotential, which consists of
the $R$--parity conserving (RpC) and violating part
\begin{equation}
  W = W_{\rm RpC} + W_{\rm RpV},
\end{equation}
where
\begin{eqnarray}
  W_{\rm RpC}&=& \epsilon_{ab} \Big[ ({\bf Y}_E)_{ij} L_i^a
  H_d^b {\bar E}_j + ({\bf Y}_D)_{ij} Q_i^{ax} H_d^b {\bar D}_{jx} 
  \nonumber \\
  &+&
  ({\bf Y}_U)_{ij} Q_i^{ax} H_u^b {\bar U}_{jx} 
  - \mu H_d^a H_u^b \Big], 
  \label{WRpC}\\
  W_{\rm RpV}&=&
  \epsilon_{ab}\left[ 
    \frac{1}{2}({\bf \Lambda}_{E^k})_{ij} L_i^a L_j^b{\bar E}_k +
    ({\bf \Lambda}_{D^k})_{ij} L_i^a Q_j^{xb} {\bar D}_{kx} \right].
  \nonumber \\
  \label{WRpV}
\end{eqnarray}
Here {\bf Y}'s are the 3$\times$3 trilinear Yukawa-like couplings, $\mu$
the bilinear Higgs coupling, and ($\bf \Lambda$) and ($\bf \kappa^i$)
are the $R$-parity violating trilinear and bilinear terms. $L$ and $Q$
denote the $SU(2)$ left-handed doublets, while $\bar E$, $\bar U$ and
$\bar D$ are the right-handed lepton, up-quark and down-quark $SU(2)$
singlets, respectively. $H_d$ and $H_u$ mean two Higgs doublets. We have
introduced color indices $x,y,z = 1,2,3$, generation indices
$i,j,k=1,2,3=e,\mu,\tau$ and the $SU(2)$ gauge indices $a,b = 1,2$.

The supersymmetry is not observed in the regime of energies accessible
to our experiments, therefore it must be broken at some
point. A~convenient method to take this fact into account is to
introduce explicit terms, which break supersymmetry in a~soft way, ie.,
they do not suffer from ultraviolet divergencies. We add them in the
form of a~scalar Lagrangian \cite{mSUGRA},
\begin{eqnarray}
  -{\cal L} &=&
  m_{H_d}^2 h_d^\dagger h_d + m_{H_u}^2 h_u^\dagger h_u
  + l^\dagger ({\bf m}_L^2) l \\ 
  &+& {l_i}^\dagger ({\bf m}_{L_i H_d}^2) h_d 
  + h_d^\dagger ({\bf m}_{H_d L_i}^2) l_i \nonumber \\
  &+& q^\dagger ({\bf m}_Q^2) q + e ({\bf m}_E^2) e^\dagger
  +  d ({\bf m}_D^2) d^\dagger + u ({\bf m}_U^2) u^\dagger \nonumber \\
  &+& \frac12 \left( 
    M_1 \tilde{B}^\dagger \tilde{B} + 
    M_2 \tilde{W_i}^\dagger \tilde{W^i} +
    M_3 \tilde{g_\alpha}^\dagger \tilde{g^\alpha} + h.c.\right )
  \nonumber \\
  &+& [ ({\bf A}_E)_{ij} l_i h_d e_j
  +    ({\bf A}_D)_{ij} q_i h_d d_j
  +    ({\bf A}_U)_{ij} q_i h_u u_j \nonumber \\
  && - B h_d h_u + \mathrm{h.c.}] \nonumber \\ 
  &+& [ ({\bf A}_{E^k})_{ij} l_i l_j e_k
  +    ({\bf A}_{D^k})_{ij} l_i q_j d_k
  +    ({\bf A}_{U^i})_{jk} u_i d_j d_k \nonumber \\
  && - D_i l_i h_u + \mathrm{ h.c.}], \nonumber
\end{eqnarray}
where the lower case letter denotes the scalar part of the respective
superfield. $M_i$ are the gaugino masses, and $\bf A$ ($B$, $D_i$) are
the soft supersymmetry breaking equivalents of the trilinear (bilinear)
couplings from the superpotential.

\subsection{Free parameters}

One of the weaknesses of the supersymmetric models is their enormous
number of free parameters, which easily may exceed 100. This lowers
significantly their predictive power. Therefore it is a custom approach
to impose certain boundary conditions, which allow in turn to
\emph{derive} other unknown parameters. One may either start with
certain known values at low energies and evaluate the values of other
parameters at higher energies, or assume, along the Grand Unified
Theories (GUT) line of thinking, common values at the unification scale
and evaluate them down. In both cases the renormalization group
equations (RGE) are used. We follow the top-down approach by assuming
the following:
\begin{itemize}
\item we introduce a common mass $m_0$ for all the scalars at the GUT
  scale $m_{\rm GUT} \approx 1.2 \times 10^{16}\GeV$,
\item we introduce a similar common mass $m_{1/2}$ for all the fermions
  at the GUT scale,
\item we set all the trilinear soft supersymmetry breaking couplings
  $\bf A$ to be proportional to the respective Yukawa couplings with a
  common factor $A_0$, ${\bf A}_i = A_0 {\bf Y}_i$ at $m_{\rm GUT}$.
\end{itemize}
The only couplings which we allow to evolve freely, are the RpV ${\bf
  \Lambda}$'s, which are set to a~common value $\Lambda_0$ at $m_Z$ and
are being modified by the RGE running only. The reason for this is, that
we want to keep them non-zero, but on the other hand their influence at
the low scale must be small. In the first part of this presentation we
will fix $\Lambda_0 = 10^{-4}$, which assures only small admixture of
the RpV interactions. Later, we discuss the impact of $\Lambda_0$ on the
results for certain set of input parameters. (For a~discussion of the
$\Lambda$'s impact on the RGE running see, eg.,
Ref.~\cite{mgozdz:art1}).

There are two remaining free parameters in the model, one describing the
ratio of the vacuum expectation values of the two Higgs doublets
$\tan\beta=v_u/v_d$, the other being the sign of the Higgs self-coupling
constant, ${\rm sgn}(\mu)$, which gives altogether only six free
parameters. In this calculations we have kept $\mu$ positive, and fixed
$\Lambda_0$, so the resultant parameter space to analyze is four
dimensional.

It is also a common practice to restrict the calculations to the third
family of quarks and leptons only, which is supposed to give the
dominant contributions. Since this issue is difficult to keep under
control, we keep dependence on all three families. Also, unlike the
Authors of Ref.~\cite{mSUGRA}, we keep the full dependence on all the RpV
couplings.

\subsection{The Higgs sector}

The procedure of finding the minimum for the scalar potential is quite
involving. The equations one has to solve in this model are given in
Ref.~\cite{mSUGRA}, however, the numerical procedure used by us differs
slightly from the one presented in the cited paper.

The goal is to find the values of $\mu$, $\kappa_i$, $B$, and $D_i$, as
well as the five vacuum expectation values $v_{u,d,1,2,3}$ of the two
Higgs bosons and three sneutrinos. The initial values for the vev's are
$v_u=v\sin\beta$, $v_d=v\cos\beta$, $v_i=0$, where $v^2=(246
\GeV)^2$. We start by setting
\begin{equation}
  \mu=\kappa_i=B=D_i=0
\end{equation}
and evaluating $g_i$, ${\bf Y}_{U,D,E}$, and ${\bf \Lambda}_{D,E}$ to
the $m_{\rm GUT}$ scale. There we impose the GUT unification conditions
and run everything down back to the $m_Z$ scale. In this first iteration
the best minimization scale for the scalar potential
\begin{equation}
  q_{\rm min}=\sqrt{[({\bf m}_U^2)_{33}]^{1/2} [({\bf m}_Q^2)_{33}]^{1/2}}
\end{equation}
is calculated. At this scale initial values of $\mu$ and $B$ are found,
according to the relations
\begin{eqnarray}
  |\mu|^2 &=& 
  \frac{m_{H_d}^2 - m_{H_u}^2 \tan^2\beta}{\tan^2\beta -1} -
  \frac{M_Z^2}{2}, \\
  B &=& \frac{\sin 2\beta}{2} (m_{H_d}^2 - m_{H_u}^2 + 2|\mu|^2).
\end{eqnarray}
Next, a~RGE run is performed from $q_{\rm min}$ to $m_Z$, but this time
the non-zero values of $\mu$ and $B$ generate non-zero values for the
$\kappa_i$ and $D_i$, providing starting point for the next
iteration. The second iteration repeats the same steps as the first one,
with the exception that now all $\mu$, $\kappa_i$, $B$, and $D_i$
contribute to the RGE running. Getting back to the (new) $q_{\rm min}$
scale, we solve for $\mu$, $B$, and $v_i$ using the loop-corrected
equations
\begin{eqnarray}
  |\mu|^2 &=& \frac{1}{\tan^2\beta -1} \biggl \{ \left [
    m_{H_d}^2 + {\bf m}_{L_iH_d}^2 \frac{v_i}{v_d} + \kappa_i^* \mu
    \frac{v_i}{v_d} \right ] \nonumber \\ 
  &-& \left[ m_{H_u}^2 + |\kappa_i|^2 -\frac{1}{2}(g^2+g_2^2)v_i^2
    -D_i~\frac{v_i}{v_u} \right] \tan^2\beta \biggr \} \nonumber \\
  &-& \frac{M_Z^2}{2},
  \label{eq:muB1}
  \\
  B &=& \frac{\sin 2\beta}{2} \Big [ 
    (m_{H_d}^2 - m_{H_u}^2 + 2|\mu|^2 + |\kappa_i|^2) \nonumber \\
    && \hskip 0.85truecm 
    + ({\bf m}_{L_iH_d}^2 + \kappa_i^*\mu)\frac{v_i}{v_d}
    -D_i~\frac{v_i}{v_u} \Big ],
    \label{eq:muB2}
\end{eqnarray}
and the so-called tadpole equations for the sneutrino vev's, which
explicitely read
%%%%%%%%%%%%%%%%
\begin{widetext}
%%%%%%%%%%%%%%%%
\begin{eqnarray}
  && v_1 [({\bf m}_L^2)_{11} + |\kappa_1|^2 + D'] 
  + v_2 [({\bf m}_L^2)_{21} + \kappa_1 \kappa_2^*] 
  + v_3 [({\bf m}_L^2)_{31} + \kappa_1 \kappa_3^*]
  = - [{\bf m}_{H_d L_1}^2 + \mu^*\kappa_1]v_d + D_1 v_u, \nonumber \\
  && v_1 [({\bf m}_L^2)_{12} + \kappa_2 \kappa_1^*] 
  + v_2 [({\bf m}_L^2)_{22} + |\kappa_2|^2 + D'] 
  + v_3 [({\bf m}_L^2)_{32} + \kappa_2 \kappa_3^*] 
  = - [{\bf m}_{H_d L_2}^2 + \mu^*\kappa_2]v_d + D_2 v_u,   
  \label{eq:tadpoles} \\
  && v_1 [({\bf m}_L^2)_{13} + \kappa_3 \kappa_1^*] 
  + v_2 [({\bf m}_L^2)_{23} + \kappa_3 \kappa_2^*] 
  + v_3 [({\bf m}_L^2)_{33} + |\kappa_3|^2 + D'] 
  = - [{\bf m}_{H_d L_3}^2 + \mu^*\kappa_3]v_d + D_3 v_u, \nonumber
\end{eqnarray}
where $D'=M_Z^2 \frac{\cos2\beta}{2} + (g^2+g_2^2)
\frac{\sin^2\beta}{2}(v^2-v_u^2-v_d^2)$. This set of three equations can
be easily solved and we get
\begin{equation}
  v_i~=\frac{\det W_i}{\det W}, \qquad i=1,2,3,
\end{equation}
where
\begin{equation}
  W~= \left(
    \begin{array}{lll}
      ({\bf m}_L^2)_{11} + |\kappa_1|^2 + D'   &
      ({\bf m}_L^2)_{21} + \kappa_1 \kappa_2^* & 
      ({\bf m}_L^2)_{31} + \kappa_1 \kappa_3^* \\
      ({\bf m}_L^2)_{12} + \kappa_2 \kappa_1^* &
      ({\bf m}_L^2)_{22} + |\kappa_2|^2 + D'   &
      ({\bf m}_L^2)_{32} + \kappa_2 \kappa_3^* \\
      ({\bf m}_L^2)_{13} + \kappa_3 \kappa_1^* &
      ({\bf m}_L^2)_{23} + \kappa_3 \kappa_2^* &
      ({\bf m}_L^2)_{33} + |\kappa_3|^2 + D'
    \end{array}
    \right ),
\end{equation}
and $W_i$ can be obtained from $W$ by replacing the $i$-th column with
\begin{equation}
  \left (
    \begin{array}{l}
     - [{\bf m}_{H_d L_1}^2 + \mu^*\kappa_1]v_d + D_1 v_u \\
     - [{\bf m}_{H_d L_2}^2 + \mu^*\kappa_2]v_d + D_2 v_u \\
     - [{\bf m}_{H_d L_3}^2 + \mu^*\kappa_3]v_d + D_3 v_u
    \end{array}
    \right ).
\end{equation}
%%%%%%%%%%%%%%
\end{widetext}
%%%%%%%%%%%%%%
The equations (\ref{eq:muB1})--(\ref{eq:tadpoles}) are solved
subsequently until convergence and self-consistency of the results is
obtained. After this procedure we add also the dominant radiative
corrections \cite{barger}, and get back to the $m_Z$ scale to obtain the
mass spectrum of the model.

For the details of the mass matrices which need to be diagonalized see
Ref.~\cite{mSUGRA}.

\section{Constraining the mass spectrum}

Not all initial parameters result in an acceptable mass spectrum. One
may impose several different constraints to test the model. The problem,
however, is in the fact that the available experimental data are in most
cases not confirmed in other experiments, not to mention that they are
very model dependent. Therefore caution is needed before such
constraints will be imposed.

\subsection{The (120--140) GeV Higgs boson}

First, we are going to check whether the recently suggested $120 \GeV <
m_{h^0} < 140 \GeV$ may be obtained within the described model. The only
additional constraints that we have used are the mass limits for
different particles, as published by the Particle Data Group in
2010. They read \cite{pdg2010}: $m_{\tilde\chi^0_1} > 46 \GeV$,
$m_{\tilde\chi^0_2} > 62 \GeV$, $m_{\tilde\chi^0_3} > 100 \GeV$,
$m_{\tilde\chi^0_4} > 116 \GeV$, $m_{\tilde\chi^\pm_1} > 94 \GeV$,
$m_{\tilde\chi^\pm_2} > 94 \GeV$, $m_{\tilde e} > 107 \GeV$,
$m_{\tilde\mu} > 94 \GeV$, $m_{\tilde\tau} > 82 \GeV$, $m_{\tilde q} >
379 \GeV$, $m_{\tilde g} > 308 \GeV$.

We have performed a~scan over the whole parameter space given by the
following ranges: $200 \GeV \le m_{0,1/2} \le 1000 \GeV$ with step of 20
GeV, $5 \le \tan\beta \le 40$ with step 5, $200 \GeV \le A_0 \le 1000
\GeV$ with step 100. During this scan we have kept $\mu>0$ and a~fixed
$\Lambda_0=10^{-4}$. For each point the mass spectrum was calculated and
confronted with the imposed constraints.

%%%%%%%%%%%%%%%%%%%%%%%%%%%%%%%%%%%%%%%%%%%%%%%%%%%%%%%%%%%%%%%%%%%%%%
% FIGS. 1-8 %%%%%%%%%%%%%%%%%%%%%%%%%%%%%%%%%%%%%%%%%%%%%%%%%%%%%%%%%%
%%%%%%%%%%%%%%%%%%%%%%%%%%%%%%%%%%%%%%%%%%%%%%%%%%%%%%%%%%%%%%%%%%%%%%

\begin{figure}
  \includegraphics[width=\columnwidth]{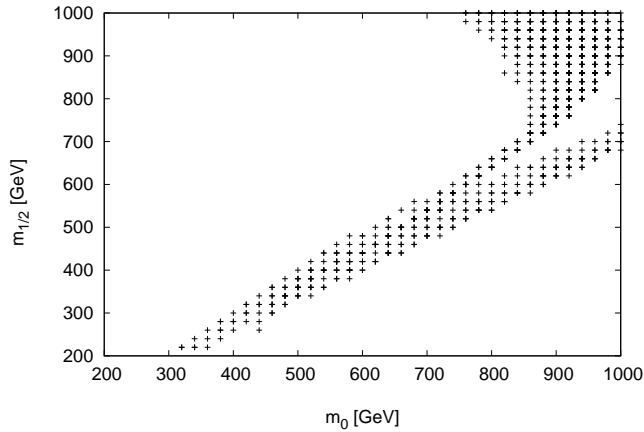}
  \caption{\label{mgozdz:fig1} Solutions for $\tan\beta=5$. For each
    point there is an $200\GeV<A<1000\GeV$ such that the mass of
    lightest Higgs boson is in the region 120--140 GeV. }
\end{figure}

\begin{figure}
  \includegraphics[width=\columnwidth]{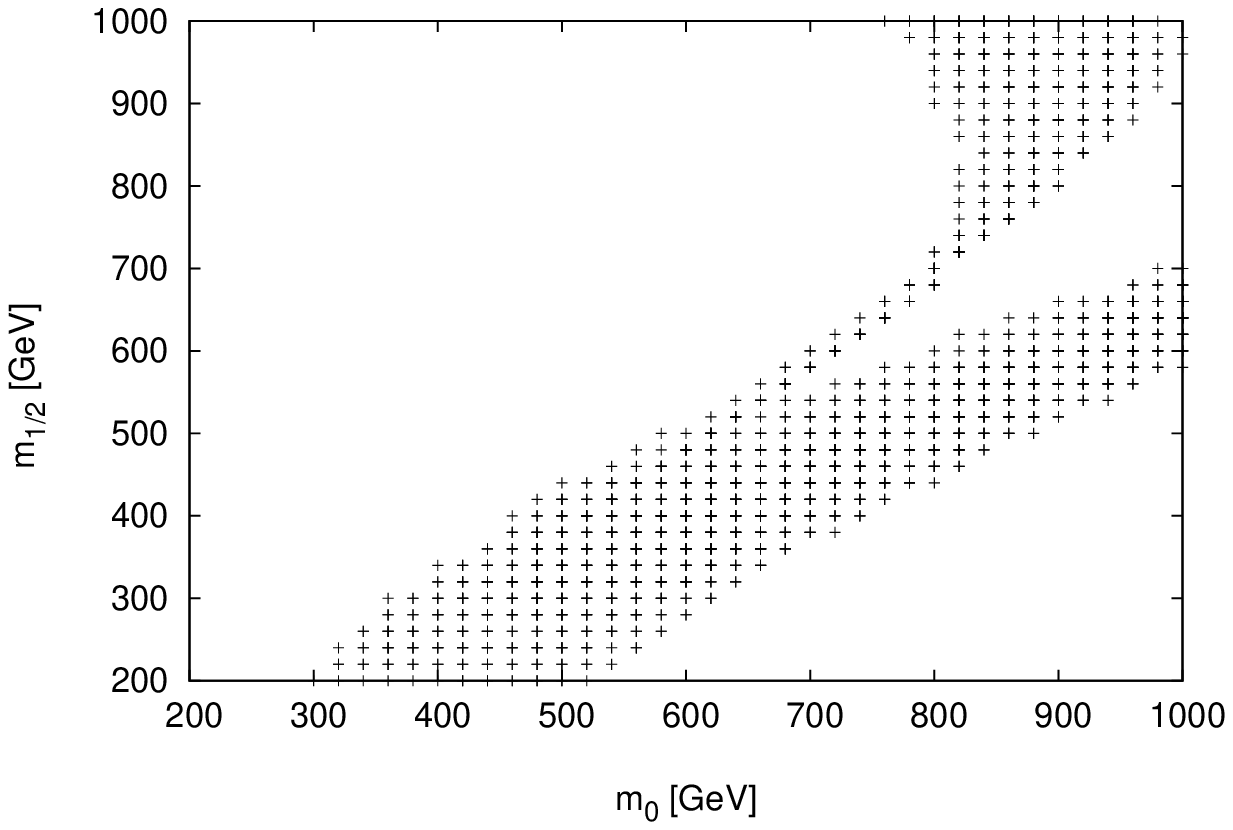}
    \caption{\label{mgozdz:fig2} Like Fig. \protect\ref{mgozdz:fig1} but
      for $\tan\beta=10$.}
\end{figure}

\begin{figure}
  \includegraphics[width=\columnwidth]{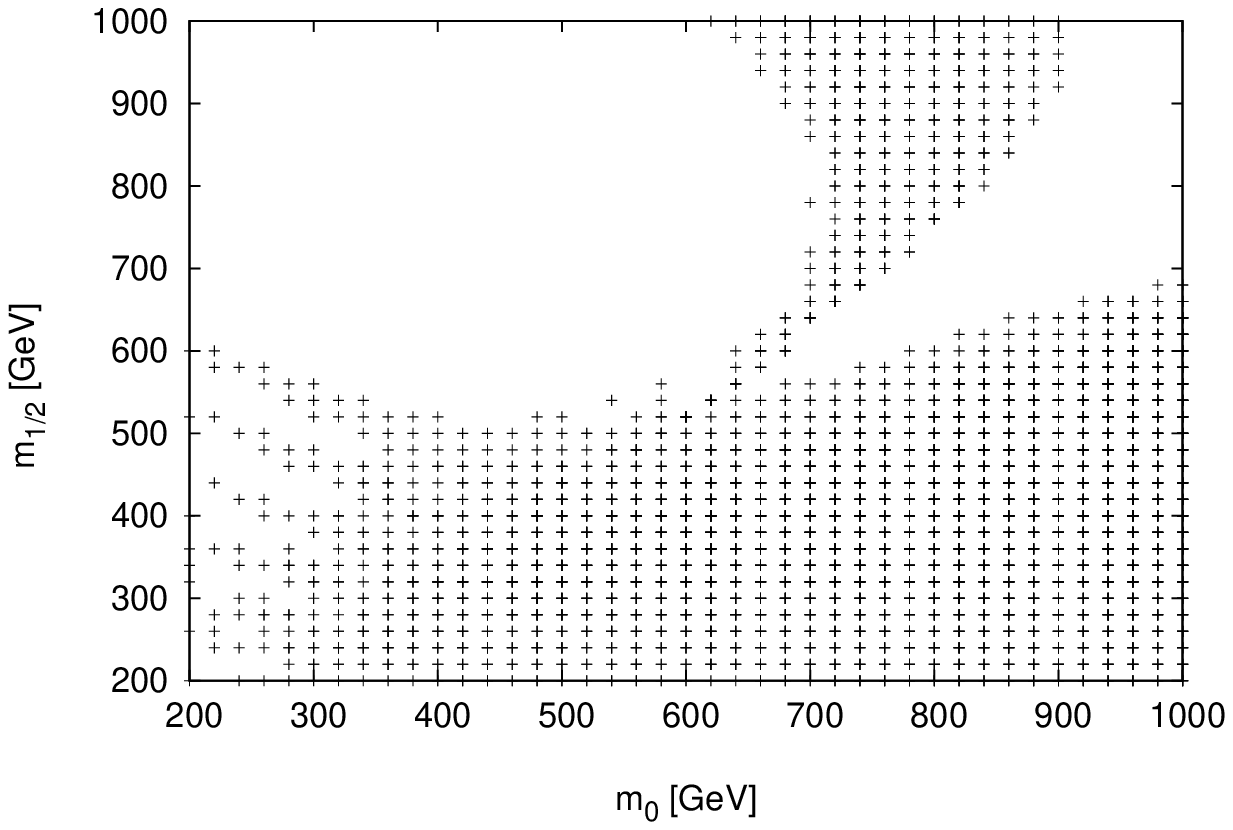}
  \caption{\label{mgozdz:fig3} Like Fig. \protect\ref{mgozdz:fig1} but
      for $\tan\beta=15$.}
\end{figure}

\begin{figure}
  \includegraphics[width=\columnwidth]{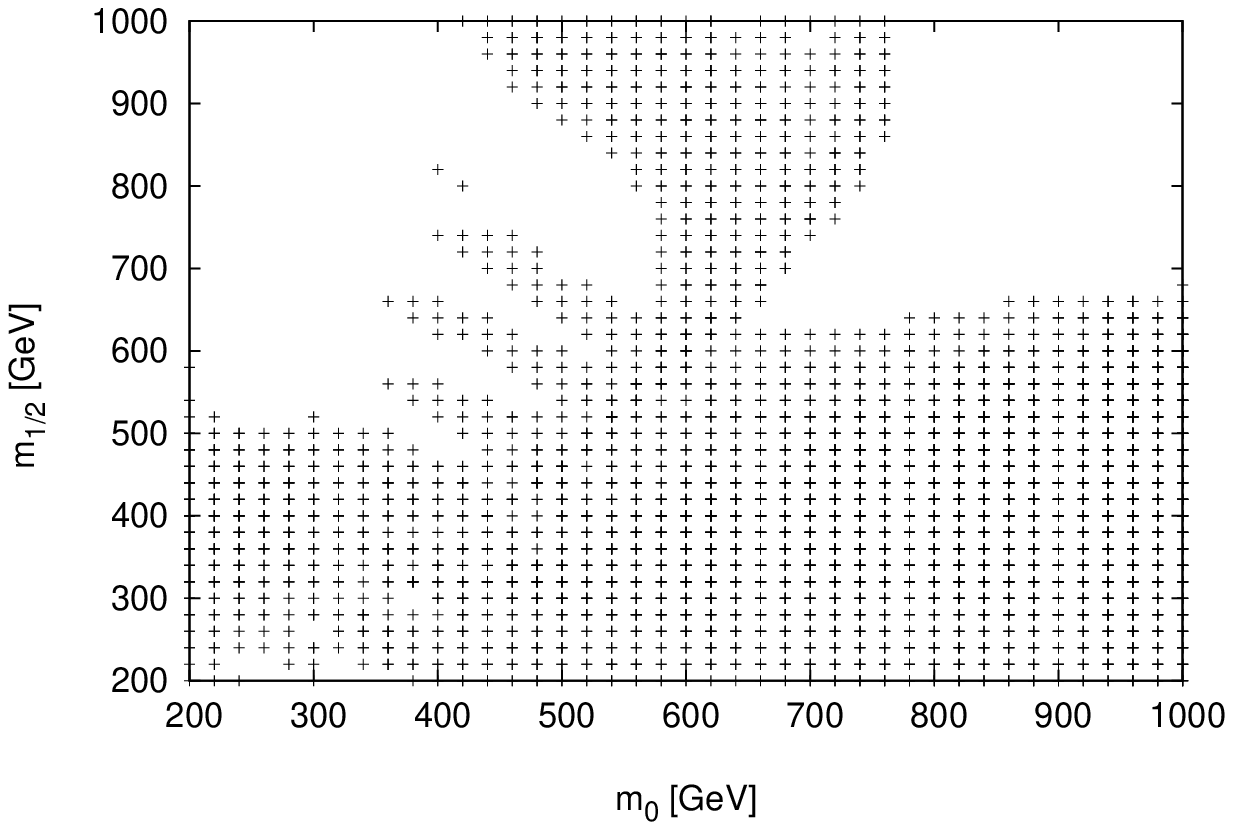}
  \caption{\label{mgozdz:fig4} Like Fig. \protect\ref{mgozdz:fig1} but
      for $\tan\beta=20$.}
\end{figure}

\begin{figure}
  \includegraphics[width=\columnwidth]{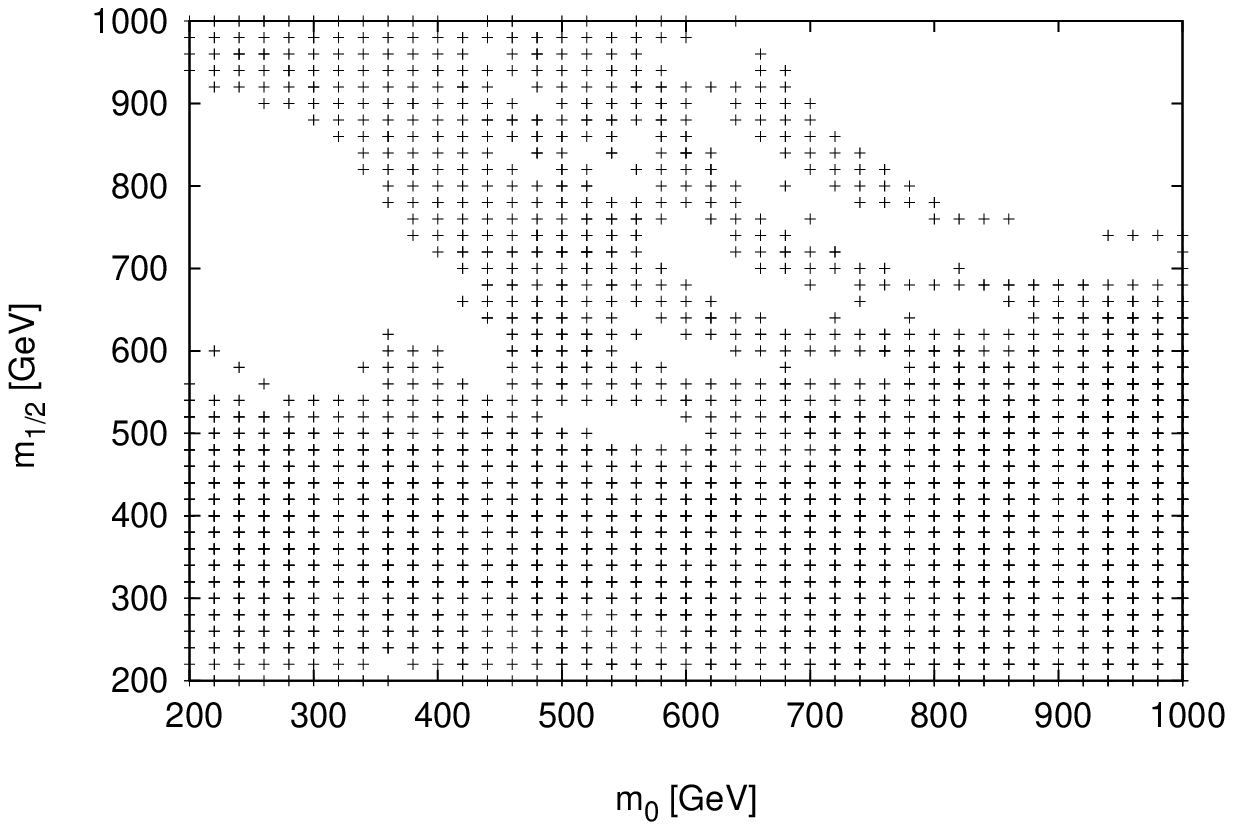}
  \caption{\label{mgozdz:fig5} Like Fig. \protect\ref{mgozdz:fig1} but
      for $\tan\beta=25$.}
\end{figure}

\begin{figure}
  \includegraphics[width=\columnwidth]{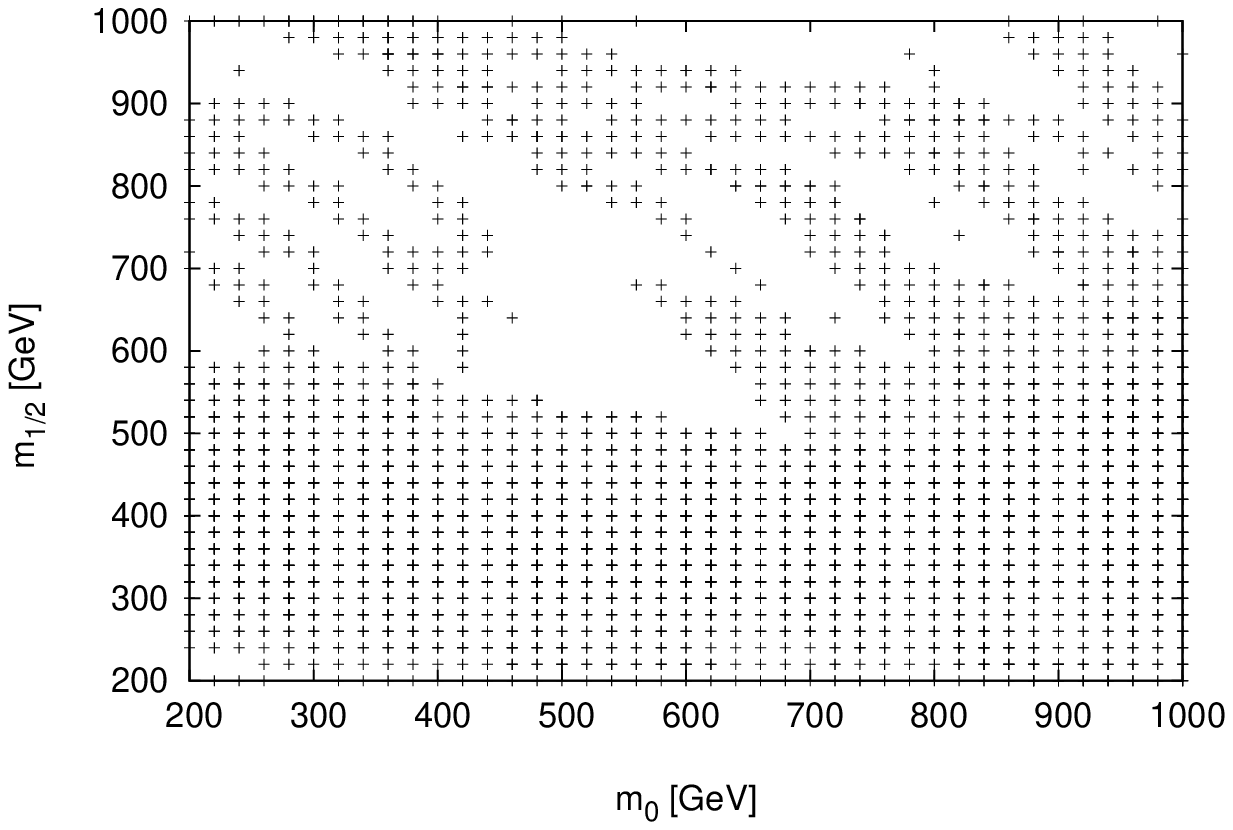}
  \caption{\label{mgozdz:fig6} Like Fig. \protect\ref{mgozdz:fig1} but
      for $\tan\beta=30$.}
\end{figure}

\begin{figure}
  \includegraphics[width=\columnwidth]{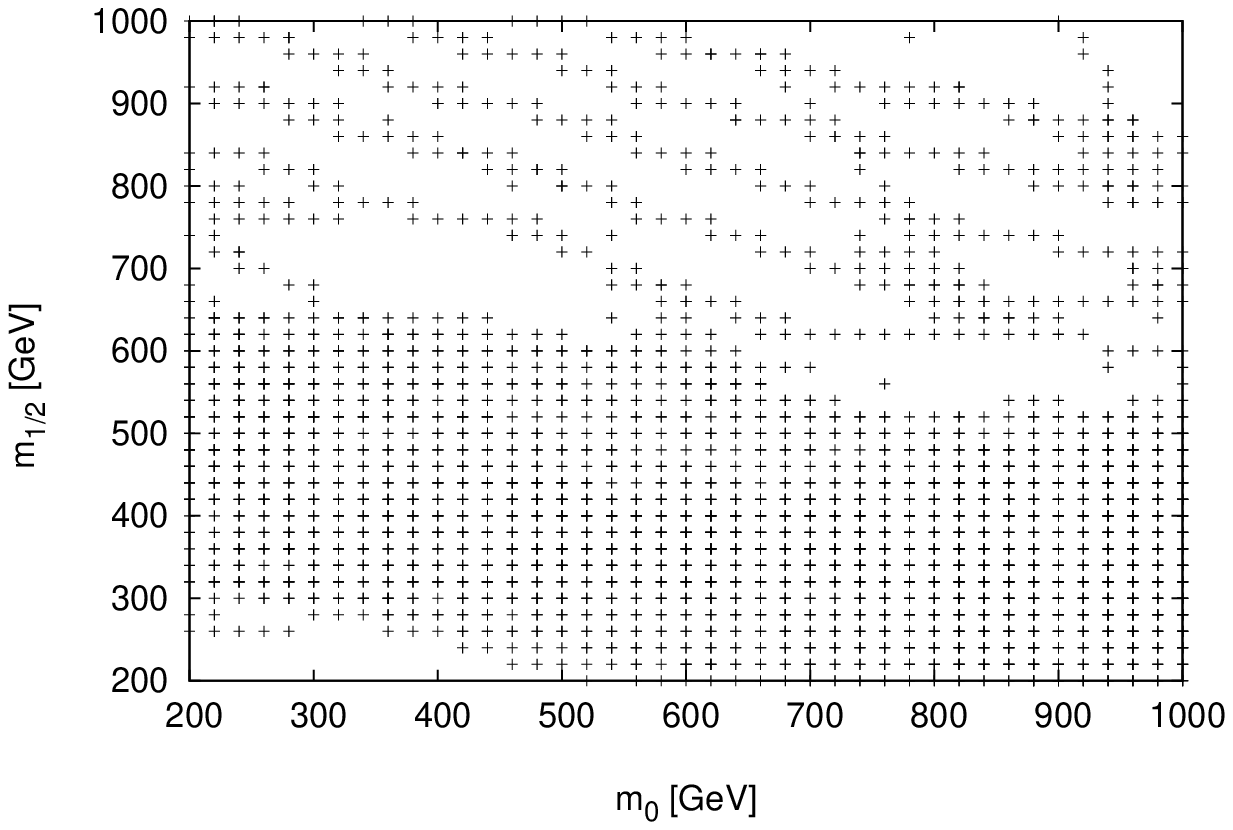}
  \caption{\label{mgozdz:fig7} Like Fig. \protect\ref{mgozdz:fig1} but
      for $\tan\beta=35$.}
\end{figure}

\begin{figure}
  \includegraphics[width=\columnwidth]{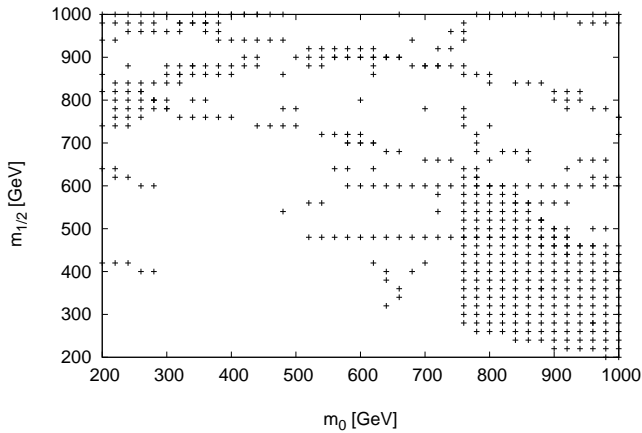}
  \caption{\label{mgozdz:fig8} Like Fig. \protect\ref{mgozdz:fig1} but
      for $\tan\beta=40$. For such high value the model breaks down and
      the results are not reliable.}
\end{figure}
%%%%%%%%%%%%%%%%%%%%%%%%%%%%%%%%%%%%%%%%%%%%%%%%%%%%%%%%%%%%%%%%%%%%%%

The results are presented in Figs.~\ref{mgozdz:fig1}--\ref{mgozdz:fig8}.
There is a~separate diagram in the $m_0-m_{1/2}$ plane for each
$\tan\beta=5,10,...,40$. Every point on these diagrams says, that for
certain values of $m_0$, $m_{1/2}$, and $\tan\beta$ there is an $A_0$
between 200 GeV and 1000 GeV for which $m_{h^0}$ is between 120 GeV and
140 GeV. Therefore Figs.~\ref{mgozdz:fig1}--\ref{mgozdz:fig8} represent
the ranges of free parameters, for which the model is compatible with
the newest experimental suggestions within error margins.

One sees that for small values of $\tan\beta \le 10$ the parameter range
is quite constrained. For higher values of $\tan\beta$ the common
fermion mass $m_{1/2}$ is preferred to be below roughly 600
GeV. A~characteristic feature of the model is presented in
Fig.~\ref{mgozdz:fig8}, ie., that it breaks down for $\tan\beta\approx
40$. This is because for such high values the bottom and tau Yukawa
couplings tend to obtain unacceptably high values during the RGE
running. Similarly, too small values of $\tan\beta \le 2$ make the top
Yukawa coupling explode.

\subsection{The 125 GeV Higgs boson}

\begin{figure*}
  \centering
  \includegraphics[width=0.8\textwidth,clip]{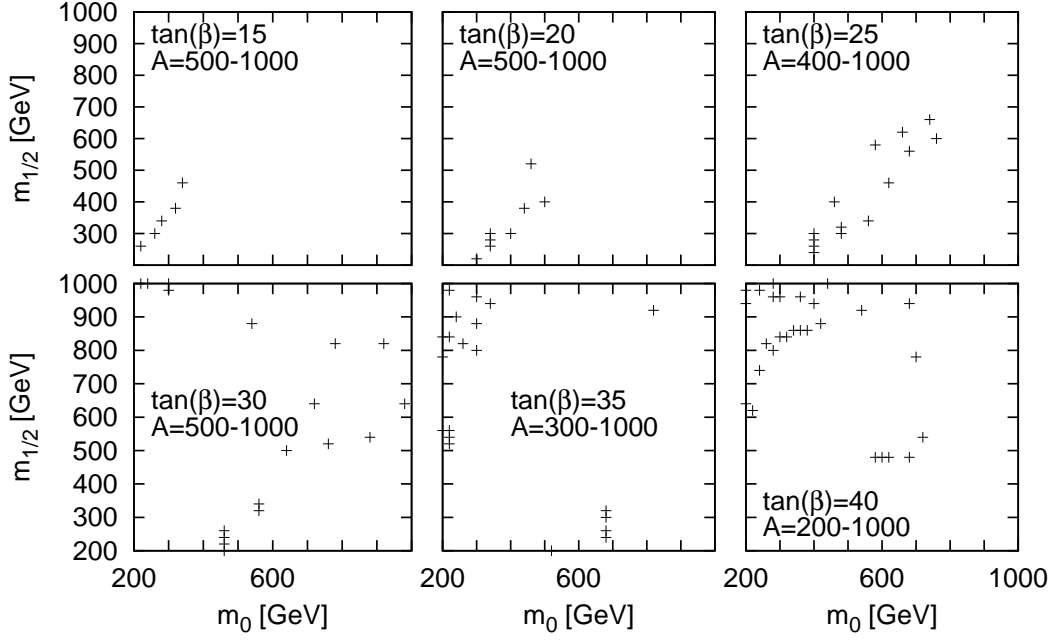}
  \caption{Parameter space of the mSUGRA model constrained by the
    condition $m_{h^0} = (125 \pm 2.5)\GeV$.}
  \label{fig:param-space}
\end{figure*}

\begin{table}
  \centering
  \caption{\label{tab:higgs1} List of free parameters of the model for
    which $m_{h^0} = (125.5\pm 2.5)\GeV$ (cf. Fig.~\ref{fig:param-space}).
    Here, $\mu>0$ and $\Lambda_0=10^{-5}$.} 
  \begin{tabular}{lllll|lllll}
    \hline\hline
    $A_0$ & $m_0$ & $m_{1/2}$ & $\tan\beta$ & $m_{h^0}$ &
    $A_0$ & $m_0$ & $m_{1/2}$ & $\tan\beta$ & $m_{h^0}$ \\
    \hline
    200   &720   & 540   & 40  & 123.7   & 700   & 260   & 300   & 15  &  125.8 \\  
    200   &700   & 780   & 40  & 125.2   & 700   & 400   & 300   & 20  & 126.1  \\ 
    200   &680   & 940   & 40  & 123.0   & 700   & 560   & 340   & 25  & 127.0  \\ 
    300   &520   & 200   & 35  & 127.4   & 700   & 300   & 1000  & 30  &  124.3 \\ 
    300   &220   & 620   & 40  & 123.2   & 700   & 540   & 880   & 30  & 123.4  \\ 
    300   &200   & 640   & 40  & 124.9   & 700   & 720   & 640   & 30  & 125.3  \\ 
    400   &460   & 200   & 30  & 126.8   & 700   & 760   & 520   & 30  & 123.4  \\ 
    400   &460   & 220   & 30  & 126.0   & 700   & 240   & 900   & 35  & 124.2  \\ 
    400   &460   & 240   & 30  & 126.4   & 700   & 300   & 880   & 35  & 126.7  \\ 
    400   &460   & 260   & 30  & 127.3   & 700   & 300   & 840   & 40  & 126.9  \\ 
    400   &680   & 240   & 35  & 124.7   & 700   & 320   & 840   & 40  & 123.2  \\ 
    400   &680   & 260   & 35  & 123.1   & 700   & 340   & 860   & 40  & 127.0  \\ 
    400   &680   & 300   & 35  & 124.3   & 700   & 360   & 860   & 40  & 124.7  \\ 
    400   &680   & 320   & 35  & 126.9   & 700   & 380   & 860   & 40  & 124.0  \\ 
    400   &280   & 800   & 40  & 126.1   & 700   & 420   & 880   & 40  & 126.5  \\ 
    400   &260   & 820   & 40  & 126.9   & 700   & 200   & 980   & 40  & 124.5  \\ 
    500   &300   & 220   & 20  & 126.9   & 700   & 240   & 980   & 40  & 125.7  \\ 
    500   &400   & 240   & 25  & 123.0   & 800   & 280   &  340  & 15 &  123.1 \\  
    500   &400   & 260   & 25  & 122.7   & 800   & 440   & 380   & 20  & 123.4  \\ 
    500   &400   & 280   & 25  & 122.8   & 800   & 620   & 460   & 25  & 124.7  \\ 
    500   &400   & 300   & 25  & 125.4   & 800   & 580   & 580   & 25  & 126.8  \\ 
    500   &560   & 320   & 30  & 122.8   & 800   & 220   & 1000  & 30 &  122.8 \\ 
    500   &560   & 340   & 30  & 125.6   & 800   & 240   & 1000  & 30 &  124.5 \\ 
    500   &200   & 780   & 35  & 126.7   & 800   & 300   & 980   & 30  & 124.3  \\ 
    500   &580   & 480   & 40  & 125.0   & 800   & 780   & 820   & 30  & 126.4  \\ 
    500   &600   & 480   & 40  & 125.2   & 800   & 880   & 540   & 30  & 126.8  \\ 
    500   &620   & 480   & 40  & 125.1   & 800   & 200   & 560   & 35  & 122.6  \\ 
    500   &680   & 480   & 40  & 125.8   & 800   & 220   & 980   & 35  & 124.5  \\ 
    500   &540   & 920   & 40  & 125.6   & 800   & 300   & 960   & 35  & 126.9  \\ 
    500   &400   & 940   & 40  & 126.2   & 800   & 340   & 940   & 35  & 126.8  \\ 
    500   &360   & 960   & 40  & 125.3   & 800   & 200   & 940   & 40  & 124.2  \\ 
    500   &280   & 1000  & 40  & 127.2   & 800   & 280   & 960   & 40  & 123.2  \\  
    600   &220   & 260   & 15  & 123.1   & 800   & 300   & 960   & 40  & 125.6  \\ 
    600   &340   & 260   & 20  & 123.4   & 800   & 440   & 1000  & 40 &  126.7 \\ 
    600   &340   & 280   & 20  & 123.6   & 900   & 320   &  380  & 15 &  125.8 \\  
    600   &340   & 300   & 20  & 125.5   & 900   & 500   & 400   & 20  & 125.0  \\ 
    600   &480   & 300   & 25  & 124.9   & 900   & 460   & 520   & 20  & 125.8  \\ 
    600   &480   & 320   & 25  & 126.1   & 900   & 680   & 560   & 25  & 123.9  \\ 
    600   &460   & 400   & 25  & 126.3   & 900   & 660   & 620   & 25  & 126.7  \\ 
    600   &640   & 500   & 30  & 126.0   & 900   & 920   & 820   & 30  & 125.6  \\ 
    600   &200   & 840   & 35  & 123.6   & 900   & 980   & 640   & 30  & 127.3  \\ 
    600   &220   & 840   & 35  & 127.3   & 900   & 220   & 520   & 35  & 125.7  \\ 
    600   &260   & 820   & 35  & 124.5   & 900   & 220   & 540   & 35  & 124.4  \\ 
    600   &300   & 800   & 35  & 127.3   & 900   & 220   & 560   & 35  & 123.3  \\ 
    600   &820   & 920   & 35  & 123.3   & 1000  &  340  &  460  & 15 &  125.3 \\ 
    600   &240   & 740   & 40  & 124.6   & 1000  &  760  &  600  & 25 &  124.9 \\ 
          &      &       &     &         & 1000  &  740  &  660  & 25 &  125.2 \\ 
    \hline\hline
  \end{tabular}
\end{table}

\begin{table}
  \centering
  \caption{\label{tab:higgs2} List of free parameters of the mSUGRA model for
    which $m_{h^0} = (125.5\pm 0.5)\GeV$. Here, $\mu>0$ and $\Lambda_0=10^{-5}$.}
  \begin{tabular}{llllll}
    \hline\hline
    point no. & $A_0$ & $m_0$ & $m_{1/2}$ & $\tan\beta$ & $m_{h^0}$ \\
    \hline
    (01) & 200   & 700   & 780 &   40 & 125.2 \\
    (02) & 500   & 360   & 960 &   40 & 125.3 \\
    (03) & 500   & 400   & 300 &   25 & 125.4 \\
    (04) & 500   & 540   & 920 &   40 & 125.6 \\
    (05) & 500   & 560   & 340 &   30 & 125.6 \\ 
    (06) & 500   & 600   & 480 &   40 & 125.2 \\
    (07) & 500   & 620   & 480 &   40 & 125.1 \\
    (08) & 500   & 680   & 480 &   40 & 125.8 \\
    (09) & 600   & 340   & 300 &   20 & 125.5 \\
    (10) & 700   & 240   & 980 &   40 & 125.7 \\
    (11) & 700   & 260   & 300 &   15 & 125.8 \\
    (12) & 700   & 720   & 640 &   30 & 125.3 \\
    (13) & 800   & 300   & 960 &   40 & 125.6 \\
    (14) & 900   & 220   & 520 &   35 & 125.7 \\
    (15) & 900   & 320   & 380 &   15 & 125.8 \\
    (16) & 900   & 460   & 520 &   20 & 125.8 \\
    (17) & 900   & 920   & 820 &   30 & 125.6 \\
    (18) & 1000  & 340   & 460 &   15 & 125.3 \\
    (19) & 1000  & 740   & 660 &   25 & 125.2 \\
    \hline\hline
  \end{tabular}
\end{table}

If we assume, according to the newest data, the lightest Higgs boson
mass to be centered around 125 GeV, the allowed parameter space shrinks
drastically. First, let us allow for a~5~GeV spread in the lightest
Higgs boson mass. On Fig.~\ref{fig:param-space} the points corresponding
to $m_{h^0} = (125 \pm 2.5) \GeV$ are presented. As a~reference, we give
the whole list in Tab.~\ref{tab:higgs1}. If we narrow our field of
interest to the, say, $(125-126)\GeV$ region only, we end up with the
parameter space listed in Tab.~\ref{tab:higgs2}. It is apparent, that
higher values of $A_0$ and $\tan\beta$ are favoured. Also, quite often
if one of the $m$'s takes smaller value, it is compensated by a high
value of the other.

It is worth to comment at this point at the recent constraint on
supersymmetry formulated by the LHC collaborations CMS and ATLAS
\cite{CMS,ATLAS}. It excludes the existence of supersymmetric particles
with masses below roughly 1 Tev. However, this conclusion has been drawn
for the simplest supersymmetric models in which, among others, the
$R$-parity is conserved, and as such do not directly apply to the model
discussed here.

\subsection{The $\Lambda_0$ dependence}

\begin{figure*}
  \centering
  \includegraphics[width=0.8\textwidth]{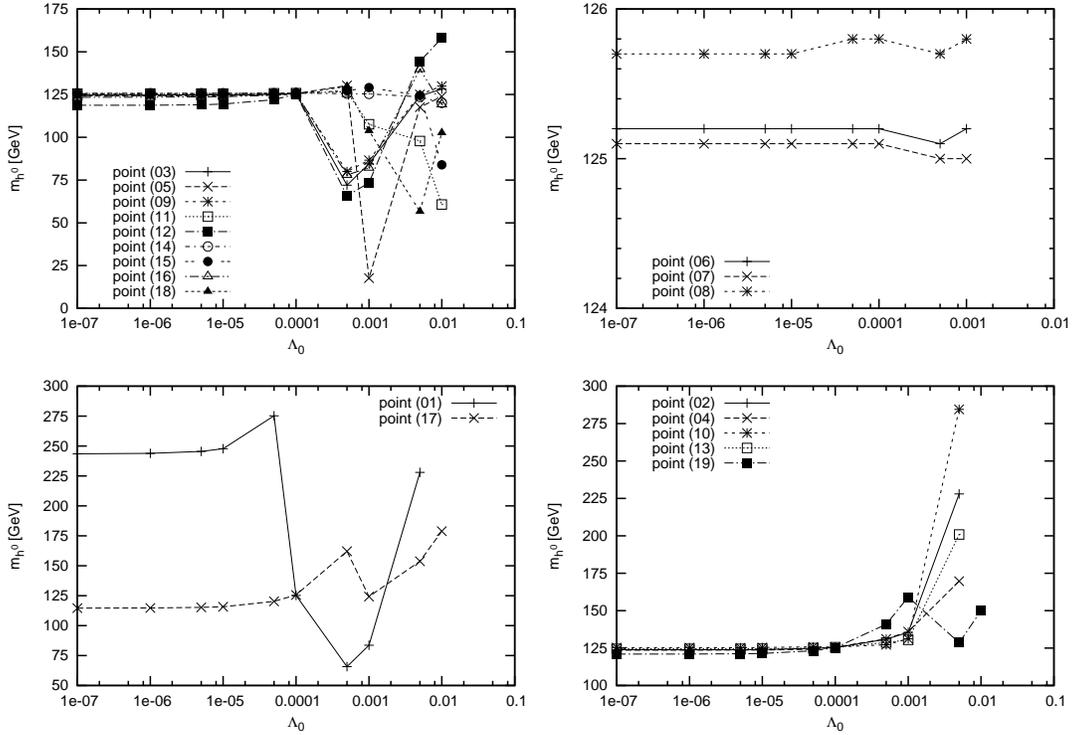}
  \caption{\label{fig:mhL} The $\Lambda_0$ dependence of the lightest
    Higgs boson mass for candidate points listed in
    Tab.~\ref{tab:higgs2}.}
\end{figure*}

All the calculations have been presented so far for a~fixed
$\Lambda_0=10^{-4}$ parameter. This is just the value for which the RpV
effects start to appear, however, their contribution is very small. For
$\Lambda_0=10^{-5}$ and below, the model essentially becomes $R$--parity
conserving. On the other hand, values of $\Lambda_0$ of the order of
$10^{-2}-10^{-1}$ have a~very big impact on the results, often throwing
the mass spectrum out of the allowed ranges. There is therefore a~rather
modest region of the $\Lambda_0$ parameters in which the model is still
physically acceptable, and at the same time the RpV contributions are
not marginal.

Let us now check what is the $\Lambda_0$ dependence of the $m_{h^0}$
mass for the 19 candidate points listed in Tab.~\ref{tab:higgs2}. We do
not expect all of them to behave in the same way under the change of the
$\Lambda_0$ parameter, especially that some of them are found in the
$\tan\beta=40$ region, which is in parts numerically unstable. The
results are presented in Fig.~\ref{fig:mhL}, where we have grouped the
points according to their functional dependence on $\Lambda_0$. On the
lower left hand side pannel two point are presented which seem to be
found by accident only, and they yield the wanted lightest Higgs boson
mass just for the $\Lambda_0=10^{-4}$ value. On the lower right hand
side and the upper left hand side pannels we present the category of
points, which converge to the correct $m_{h^0}$ value for decresing
$\Lambda_0$. These points would be the sought solutions in the
$R$--parity conserving model, and also in the RpV case presented here
for fine-tuned values of $\Lambda_0$. We see that deviations from the
$m_{h^0}\approx 125\GeV$ may be substantial for $\Lambda_0$ greater than
$\mathrm{few}\times 10^{-4}$, with general tendency to increase (lower
right hand side pannel) or decrease/oscillate (upper left hand side
pannel). The last, upper right hand side pannel shows three points which
very weakly depend on the changing of $\Lambda_0$. The solutions
obtained form them are stable, regardles in the RpC and RpV
regime. These points, surprisingly, also have $\tan\beta=40$.

\section{Conclusions}

It was interesting to check, that for the typical minimal supergravity
model with broken $R$-parity there is a~quite wide parameter space, for
which the mass of the lightest Higgs boson is compatible with the recent
Tevatron observations. However, if one refines the constraints using the
LHC--7 results, the parameter space shrinks drastically to a~set of
points roughly given in Tab.~\ref{tab:higgs2}. In this communication we
have used a~very modest set of constraints on the low-energy spectrum,
keeping only the most obvious ones. A~more detailed analysis containing
a~discussion on the Higgs and higssino contributions to the neutrino
magnetic moment and 1-loop neutrino masses will be given elsewhere.

We may conclude that, in the way presented above, we have found 17 good
candidate points in the RpV mSUGRA model which result in a~physically
acceptable mass spectrum and are at the same time compatible with the
newest Higgs boson searches. There are many more points, like the one
numbered (1) and (17), which would also give the desired mass spectrum,
but for which a~specific fine--tuning of the parametes is necessary.

This project has been financed by the Polish National Science Centre
under the decission number DEC-2011/01/B/ST2/05932. Also the support of
the European Community -- Research Infrastructure Action under the FP7
``Capacities'' Specific Programme is greatly acknowledged.

%%%%%%%%%%%%%%%%%%%%%%%%%%%%%%%%%%%%%%%%%%%%%%%%%%%%%%%%%%%%%%%%%%%%%%

%%%%%%%%%%%%%%%%%%%%%%%%%%%%%%%%%%%%%%%%%%%%%%%%%%%%%%%%%%%%%%%%%%%%%%

\end{document}